\documentclass{aa}  
\usepackage{amsmath}
\usepackage{amssymb}
\usepackage{amstext}
\renewcommand{\vec}[1]{\ensuremath{\mathbf{#1}}}
\usepackage{natbib}
\bibpunct{(}{)}{;}{a}{}{,}
\usepackage{graphicx}
\usepackage{txfonts}
\begin{document}
 \title{Broadband Fizeau Interferometers for Astrophysics}


   \author{Siddharth S. Malu
          \inst{1}
          \and
          Peter T. Timbie\inst{2}
          }

   \institute{Raman Research Institute, Sadashivanagar, Bangalore 560 080, India. \email{siddharth@rri.res.in}
         \and
         University of Wisconsin-Madison, Madison WI 53706, USA. \email{pttimbie@wisc.edu}
             }

   \date{Received ; accepted }


\abstract{Measurements of the 2.7 K cosmic microwave background (CMB) radiation now provide the most stringent constraints on cosmological models. The power spectra of the temperature anisotropies and the $E$-mode polarization of the CMB are explained well by the inflationary paradigm.  The next generation of CMB experiments aim at providing the most direct evidence for inflation through the detection of $B$-modes in the CMB polarization, presumed to have been caused by gravitational waves generated during the inflationary epoch around $10^{-34}$s. The $B$-mode polarization signals are very small ($\leq$10$^{-8}$K) compared with the temperature anisotropies ($\sim 10^{-4}$K).  Systematic effects in CMB telescopes can cause leakage from temperature anisotropy into polarization.  Bolometric interferometry (BI)  is a novel approach to measuring this small signal with lower leakage.}{If BI can be made to work over wide bandwidth ($\sim20-30\%$) it can provide similar sensitivity to imagers. Subdividing the frequency passband of a Fizeau interferometer would mitigate the problem of `fringe smearing.'  Furthermore, the approach should allow simultaneous measurements in image space and visibility space.}
           {For subdividing the frequency passsband (`sub-band splitting' henceforth), we write an expression for the output from every
             baseline at every detector in the focal plane as a sum of visibilities in different
     frequency sub-bands. For operating the interferometer simultaneously as an imager, we write the
     output as two integrals over the sky and the focal plane, with
     all the phase differences accounted for.}{
The sub-band splitting method described here is general and
can be applied to broad-band Fizeau interferometers across the electromagnetic spectrum.  Applications
to CMB measurements and to long-baseline optical interferometry are promising.}
{}

   \maketitle
%

\section{Introduction}
\label{intro}
Interferometry has a long history for astronomical measurements at radio, millimeter, submillimeter, IR and optical wavelengths.  In 1868 Hyppolyte Fizeau (Fizeau 1868)  described how the diameters of stars could be measured by optical interferometry.   He proposed `masking' the aperture of a telescope to create interference fringes in the focal plane, similar to a Young's double slit interference experiment. Today  aperture masks are used  to overcome atmospheric `seeing' effects to reach the diffraction limit of single aperture telescopes \citep{2000PASP..112..555T}.
For even greater angular resolution, beam combination from widely separated apertures is used in long baseline optical interferometers.    Michelson used this technique to measure, for the first time, the diameters of stars.   He used flat mirrors to reflect the beams from separated apertures into a single telescope, which acted as the beam combiner.

This type of beam combination, in which beams are combined in the image plane of a telescope, is called `Fizeau interferometry,' or `adding interferometry.'   An alternate approach involves combining the beams in the pupil plane of a telescope and is called `Michelson interferometry' after the technique used in the Michelson-Morley experiment \citep{2003RPPh...66..789M}.
Both types of combiners are used in long-baseline optical interferometers \citep{2000plbs.conf...31T}.
Conventional radio interferometers can also be thought of as Michelson interferometers.  They  typically mix the RF signals from each antenna in an array to lower frequencies (heterodyning) and interfere (multiply) them electronically one pair at a time with either an analog or digital correlator to measure visibilities.  This approach is sometimes called `multiplying' interferometry.  These  techniques are widely used for radio wavelengths to sub-mm wavelengths.   As the number of antennas, $N$,  increases the number of correlations to be performed grows as $N(N-1)/2$.  Although radio correlators are improving rapidly, they are currently limited to combining signals from about 100 antennas and bandwidths of a few GHz \citep{lawrence2008}.  

In this paper we describe a Fizeau `adding' interferometer that overcomes the bandwidth and large-$N$ limitations of heterodyne interferometers.  The instrument is optimized for precision measurements of the temperature and polarization anisotropy of the CMB and is sometimes called a `bolometric interferometer' \citep{2006NewAR..50..999T,2008SPIE,  2008arXiv0806.0380C, 2008arXiv0807.0438H, 2010EAS....40..399H}.  Interferometers are less sensitive to some kinds of systematic effects found in imaging instruments \citep{2007PhRvD..75h3517B}.  The technique can be used at any wavelength.

In particular, we focus on an approach to broadening the bandwidth of a Fizeau interferometer.  Spectral resolution is not required for many applications where the source has a continuum spectrum (such as the CMB) and signal averaging over broad bands is required to uncover faint signals.
As one increases the bandwidth the different wavelengths within the passband will smear out the interference fringes and reduce the sensitivity of the instrument.  Furthermore, this `fringe washing' reduces the resolution of the interferometer in the u-v plane.   In this paper we present a simple and powerful technique, which we call `sub-band splitting' to overcome the limitations caused by large bandwidths in Fizeau interferometers.  We discuss how this approach can improve parameter estimation in the specific case of observations of the CMB, where sensitivity and u-v space resolution are critical for constraining CMB power spectra.  

We begin by motivating the need for adding interferometry in
\S\ref{context}, followed by the reason sub-band splitting is
required. \S\ref{section4} develops the formalism to describe how
visibilities are measured by a Fizeau combiner (Appendix \ref{fizeauimage} points out how this arrangement can be used
as an imager). \S\ref{section3} describes in detail the technique for solving for visibilities in sub-bands.

\section{Cosmological Context: The CMB and Adding Interferometry}
\label{context}
We now have a `standard model' of cosmology
in which 
the inflationary paradigm
describes many aspects of the universe accurately through well-quantified
parameters. Precision measurements of the CMB are the
most powerful probes for determining these parameters.  In particular, measurements of the angular power spectrum of the CMB temperature and $E$-mode polarization anisotropy have yielded a wealth of information about the early universe
\citep[see][for~example]{2009ApJS..180..330K}.    However, 
inflation is driven by physics that we do not currently
understand \citep{2009AIPC.1141...10B}. The most direct way to probe
inflation is through the so-called $B$-modes in CMB polarization.   The amplitude of the $B$-mode signal is directly related to the energy scale of the particle interactions that occurred during inflation.  However, the energy scale of inflation is not known. 
Recent measurements  \citep{2010ApJ...716.1040G, 2010ApJ...711.1123C},  are approaching the level at which $B$-modes are expected to appear in models in which inflation occurred at the GUT scale.   However,  inflation may have involved lower-energy interactions;  there is no lower bound to the amplitude of the $B$-modes.

Current \citep{2010ApJ...711.1141T, 2009ApJ...692.1221H} and planned CMB instruments \citep{2009arXiv0906.1188B} all use some type of imaging technique.  With focal-plane arrays of hundreds of background-limited detectors they are capable of detecting the $B$-mode signals predicted in the most optimistic models, at the level of $\sim 10^{-8}$~K.   Systematic effects have been extensively studied for imaging polarimeters in the context of CMB measurements and appear to be controllable at this level as well  \citep{2003PhRvD..67d3004H, 2006astro.ph..4101B}.  However, at some level all instruments can `mix' the relatively large temperature anisotropy and $E$-mode polarization signals into $B$-modes.  

Different systematic effects are found in interferometers.  It is for this reason that heterodyne interferometers have been used for many years to study the CMB
temperature and polarization power spectra and the Sunyaev-Zel'dovich
effect.  These instruments multiply together the RF signals from all possible pairs of antennas (baselines) in an array to measure a set of visibilities, which are related to the sky image through a Fourier transform  \citep{Rohlfs&Wilson}.  In fact, the first detection of CMB $E$-mode polarization was
made by a multiplying interferometer: DASI \citep{2002Natur.420..772K}.  DASI had 13 single-mode antennas and performed pairwise correlation of signals across the Ka band, from 26 - 36 GHz.  

Several groups have studied the possibility of building a new generation of
mm-wave interferometers specifically to search for the small polarization
signals in the CMB \citep{2006NewAR..50..999T}.  Compared to existing mm-wave interferometers, these new
instruments would have to do the following: 1) collect more modes of radiation from the
sky by including more antennas ($>$ 100 ); 2) operate with broader bandwidth ($\sim 25 \%$, corresponding to $>10$~GHz), and 3) operate over a broader range of center frequencies, at least up
to 90 GHz, to be able to detect and reject astrophysical foreground
sources by their spectral signatures.  Because pairwise correlation requires multiplying $N(N-1)/2$ signals together, conventional heterodyne interferometers face a significant challenge to increasing the number of
antennas.  


On the other hand, adding interferometers have the ability to
correlate large numbers of inputs over wide bands.    Here we present an adding
interferometer based on a Fizeau beam combiner.   Combined with the
optimal phase-shifting scheme described in
\cite{2009A&A...497..963C}, this is a promising approach for measuring the $B$-mode
polarization.  The technique is compatible with either coherent receivers (amplifiers) or incoherent detectors (bolometers).


In a 2-element adding interferometer the electric field wavefronts from
both antennas are added and then squared in a detector \citep{Rohlfs&Wilson}.
( See Fig. \ref{fig:addvsmult}.)
The result is a constant term proportional to the intensity plus an
interference term.  The constant term is an offset that is removed by
phase-modulating one of the signals.  Phase-sensitive detection at the
modulation frequency recovers both the in-phase and quadrature-phase
interference terms and reduces susceptibility to low-frequency drifts
($1/f$ noise) in the detector and readout electronics. The adding
interferometer recovers the same visibilities as a multiplying
interferometer:

\begin{equation}
\label{visibility}
V(\vec{u})=\int\int I(\vec{\hat{n}},\nu)G(\vec{\hat{n}})e^{i2\pi \vec{u}\cdot\vec{\hat{n}}}J(\nu)d\nu d\vec{\hat{n}}
\end{equation}

\noindent where $I(\vec{\hat{n}},\nu)\propto |E_0(\vec{\hat{n}},\nu)|^2$ is the incident intensity and $E_0(\vec{\hat{n}},\nu)$ is the incident electric field as a function of position of the source on the sky, $\vec{\hat{n}}$, and frequency, $\nu$. $G(\vec{\hat{n}})$ is the primary beam (power) pattern of the antennas 
(assumed identical and for simplicity assumed independent of frequency). $\vec{u}$ is the vector between the centers of the antenna apertures, measured in wavelengths, and has components u and v.  The frequency bandpass of the instrument is $J(\nu)$.

To combine signals from $N>2$ antennas, we use a Fizeau beam combiner, a type of `image plane' combiner \citep{2000plbs.conf...31T}.  This technique is analogous to the simplest interferometer in 1-d: the Young's
double (or multiple)-slit interferometer  (Fig.(\ref{youngs}).   Schematics of Fizeau combination are shown in
Fig. \ref{fizeau1d}).  The beams from the apertures can directly illuminate an array of detectors, or, more typically, they pass through a lens or telescope first to reduce the size of the instrument.  
While Fizeau combining is
well-known, we stress here the fact that there are two types of path differences
(and therefore phase differences) for rays traveling from a source to a detector:  one path difference occurs  \emph{outside} the instrument, and the other \emph{inside} the
instrument.  Compare this to a conventional interferometer (also shown in
1-d, though the extension to 2-d is straightforward) as shown in
Fig.(\ref{youngs}), where rays only undergo a phase difference before they enter the antennas.  (Note that long-baseline optical interferometers usually include a `delay line'  between the apertures and the beam combiner.  This element introduces an equal path length for all rays entering an aperture.  In contrast, the internal path differences we are concerned with are different for each pixel in the focal plane.)

Let us explore what this combination of path differences achieves. We start by noting that the `external' phase differences, which are present in any interferometer, are the reason that the visibility function 
is a Fourier transform of the image on the sky. The visibility measured by a single baseline essentially 
selects one Fourier mode from the image. In the
Fizeau system, we have an additional set of phase differences. Without
loss of generality, we may assign a negative sign to the phases introduced
inside the instrument. Now, if we sum over both the phases, we get a
Fourier transform followed by an inverse Fourier transform - but this
is the image itself! Thus, Fizeau combination enables imaging in
an interferometer. The image formed in the focal plane of the Fizeau combiner is equivalent to the `dirty image' measured by conventional radio interferometers.  This topic is discussed later in this paper in Appendix~\ref{fizeauimage}.



In addition, we show that, given enough detectors on the focal
plane, we can extract some spectral information from the interference fringes and determine the visibilities in several sub-bands. 
The Fizeau system enables extraction of spectral information via
geometry, without additional components like filters.  

Without spectral information an interferometer with a large bandwidth suffers from a large radial width of each
pixel in the u-v plane. This is shown in Fig.(\ref{302}).  
Let $\nu$ be the center
frequency and $\Delta\nu$ the bandwidth. Then, a baseline of length
$B$ will measure the CMB power spectrum over a band centered on spherical harmonic number
\begin{equation}
\label{fizeau01}
\ell = \frac{\pi B}{\lambda} = \frac{\pi\nu B}{c} = \pi \sqrt{{\rm u}^2+{\rm v}^2}
\end{equation}
where the width in $\ell$-space is
\begin{equation}
\label{fizeau102}
\Delta\ell = \frac{\pi\Delta\nu B}{c}.
\end{equation}
As mentioned above, the additional spectral information that is available to us can be used to sub-divide the band in the u-v plane.   We discuss
this aspect in detail in \S\ref{section3}.


These advantages are studied in the context of two novel millimeter-wave interferometers:  MBI (the  Millimeter-wave Bolometric Interferometer, 
 \citep{2006NewAR..50..999T}) and QUBIC \citep{2010EAS....40..399H}.
In these systems the apertures are replaced by an array of
back-to-back single-mode horn antennas (see
Fig.(\ref{fig:adding_block})).  The inward--facing horns illuminate a
Cassegrain telescope that combines the beams at the focal (image)
plane, which is tiled by an array of bolometers whose dimensions are small
compared to the overall dimensions of the focal plane.   Between the outward and inward facing horns are electronic  phase modulators which can be operated independently. They are used to modulate the interference fringes that appear on the focal plane in such a way that fringes caused by different baselines can be distinguished from each other \citep{2009A&A...497..963C, 2009MNRAS.393..531H}.   In principle this technique could work with multimode horns or antennas as well.

\section{The Fizeau combiner output and its relation to visibilities}
\label{section4}
\begin{figure}
\centering
\includegraphics[width = 7.5cm]{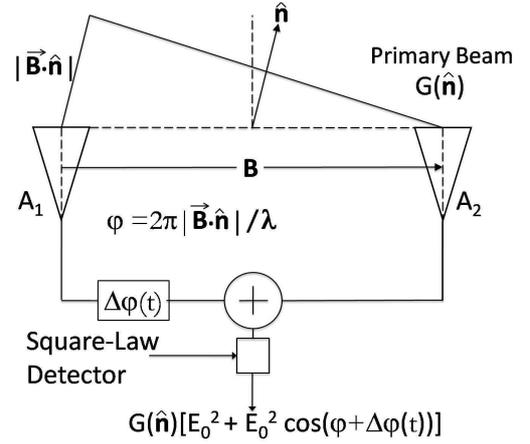}
\caption{\small\small Schematic of an adding interferometer with $N = 2$ antennas.  At antenna $A_2$ the electric
field is $E_0$, and at $A_1$ it is $E_0\exp{i\varphi}$, where $\varphi=2\pi\vec{u}\cdot\vec{\hat{n}}$ and $|\vec{u}| = |\vec{B}|$/$\lambda$.  $|\vec{B}|$ is the length of the
baseline, and $(\vec{B}\cdot\vec{\hat{n}})$/$|\vec{B}|$ is the angle of the source with respect to the
symmetry axis of the baseline. (For simplicity consider only one wavelength, $\lambda$, and ignore time dependent factors.)
In a multiplying interferometer the in-phase output of the correlator
is proportional to $ E_0^2\cos\varphi$. For the adding interferometer,
the output is proportional to $ E_0^2 + E_0^2\cos(\varphi+\Delta\varphi(t))$. Modulation of $\Delta\varphi(t)$ allows
the recovery of the interference term, $E_0^2\cos\varphi$, which is
proportional to the visibility of the baseline.}
\label{fig:addvsmult}
\end{figure}

\begin{figure}
\centering
\includegraphics[width=9.5cm,angle=0]{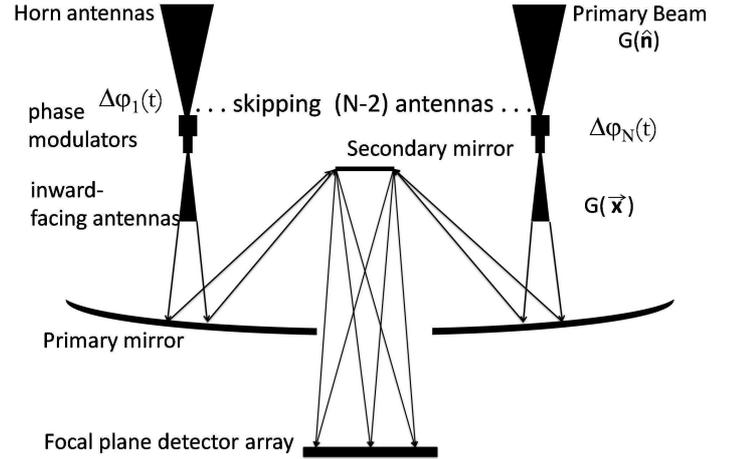}
\caption{\small\small Block diagram of an adding interferometer with $N>2$.  Each phase shifter is modulated in a
sequence that allows recovery of the interference terms (visibilities)
by phase-sensitive detection at the detectors.  The signals are mixed
in the beam combiner and detected.
The beam combiner can be implemented either using guided waves (e.g.  in a Butler
combiner) or quasioptically (Fizeau combiner), as above.  The top triangles represent corrugated conical horn antennas. For the case of an interferometer using coherent receivers, amplifiers and/or mixers could be placed before the beam combiner.}
\label{fig:adding_block}
\end{figure}

%
In this section, we study the output of the adding interferometer and
its dependence on instrument parameters: number of detectors on the
focal plane, number of antennas, etc. and relate this output to
visibility from an interferometer, in order to describe bandwidth
splitting in the following section. In what follows, we denote the output at the detectors as
$\mathcal{O}$. A simple adding interferometer with two
antennas/apertures (single baseline) is shown in Fig.(\ref{fig:addvsmult}). A
generalized Fizeau adding interferometer is shown in Fig.(\ref{fizeau1d}).

\subsection{Simple Interferometer}
\label{simple_interferometer}
In an adding interferometer, electric fields are added and then
squared at the detector. If $\vec{E}_1(\vec{\hat{n}},\nu)$ and $\vec{E}_2(\vec{\hat{n}},\nu)$
are electric fields incident at antennas 1 and 2 respectively, then, denoting the primary
beam of the outward-facing antennas by $G(\vec{\hat{n}})$, an adding interferometer will detect
\begin{equation}
\label{zero1}
G(\vec{\hat{n}})|\vec{E}_1(\vec{\hat{n}},\nu) + \vec{E}_2(\vec{\hat{n}},\nu)|^2
\end{equation}
from a certain direction $\vec{\hat{n}}$. Now $\vec{E}_1(\vec{\hat{n}},\nu)$ and $\vec{E}_2(\vec{\hat{n}},\nu)$
differ only by a phase factor
$\varphi=2\pi\vec{B}\cdot\vec{\hat{n}}$/$\lambda$, so that we can
write the electric fields as $\vec{E}_0(\vec{\hat{n}},\nu)$ and
$\vec{E}_0(\vec{\hat{n}},\nu)\exp(i\varphi)$, and the output at a single frequency for the
simple arrangement shown in Fig.(\ref{fig:addvsmult}) is
\begin{eqnarray}
&&G(\vec{\hat{n}})|\vec{E}_0(\vec{\hat{n}},\nu) + \vec{E}_0(\vec{\hat{n}},\nu)\exp(i\varphi)|^2\nonumber\\
&=&G(\vec{\hat{n}})|\vec{E}_{0}(\vec{\hat{n}},\nu)|^{2}\left(1+\cos\varphi\right)\label{some1}\\
&\equiv& G(\vec{\hat{n}})|\vec{E}_{0}(\vec{\hat{n}}),\nu|^{2}\left(1+\Re[\exp(i\varphi)]\right)\label{2nd}
\end{eqnarray}
$E_{0}^{2}(\vec{\hat{n}},\nu) \propto I(\vec{\hat{n}},\nu)$ where $I(\vec{\hat{n}},\nu)$, the intensity,
is a linear combination of Stokes parameters
\citep[see][e.g.]{2006APh....25..151L}. 
To isolate the interference term, a time--varying phase modulation is
applied, as shown in Fig.(\ref{fig:addvsmult}). Then, the demodulated output is given by
\begin{equation}
\label{seven2}
 G\left(\vec{\hat{n}}\right)I(\vec{\hat{n}},\nu)\cos\varphi\equiv G\left(\vec{\hat{n}}\right)I(\vec{\hat{n}},\nu)\Re[\exp(i\varphi)]
\end{equation}
where
$\varphi=2\pi\vec{B}\cdot\vec{\hat{n}}$/$\lambda=2\pi\vec{u}\cdot\vec{\hat{n}}$. 
\subsection{Fizeau Interferometer}
\label{fizeau_interferometer}
The general arrangement of a Fizeau interferometer is shown in
Fig.(\ref{fizeau1d}). Let $\vec{B}_{k}$ be the baseline formed by antennas at
$\vec{A}_p$ and $\vec{A}_l$, such that
$\vec{B}_{k}=\vec{A}_p-\vec{A}_l$ ($k\in\left[1\ldots N(N-1)/2\right]$
and $p,l\in\left[1\ldots N\right]$), and $\vec{\hat{n}}$ is a direction in
the sky, 
as shown in Fig.(\ref{fizeau1d}). Then, $\vec{u}_{k}=\vec{B}_{k}$/$\lambda$, so that the external path
difference is $(\vec{A}_p-\vec{A}_l)\cdot\vec{\hat{n}}=\vec{B}_{k}\cdot\vec{\hat{n}}$, as shown in
Fig.~(\ref{fizeau1d}). 

The difference between this arrangement and that
of the simple adding interferometer in Fig.(\ref{fig:addvsmult}) is the
set of internal phase differences introduced inside the instrument due
to the fact that there is more than one detector, and the geometry of
the arrangement. Let $\vec{x}_j$ denote the position of the $j$th detector on the focal
plane. Then, the internal path difference between the rays from antennas $p$ and $l$, (which form baseline
 $k$), is $x_{jk}(\vec{x})=\vec{x}_j\cdot\vec{B}_{k}$ as shown in
Fig.(\ref{fizeau1d}). Thus, the intensity at the detector at $\vec{x}_j$
contributed by baseline $k$ is
\begin{equation}
\label{seven3}
G\left(\vec{\hat{n}}\right)I(\vec{\hat{n}},\nu)\Re[\exp{i(\varphi_{k} + \frac{2\pi}{\lambda}x_{jk})}].
\end{equation}
\noindent where
$\varphi_k=2\pi\vec{B}_{k}\cdot\vec{\hat{n}}$/$\lambda=2\pi\vec{u}_k\cdot\vec{\hat{n}}$.

If we denote the output at the $j$th detector from the $k$th baseline
as $\mathcal{O}_{jk}$, and integrate over all directions in the sky and the bandwidth $\Delta\nu$, we get
\begin{equation}
\label{seven4}
\mathcal{O}_{jk} = \int\int G\left(\vec{\hat{n}}\right)I(\vec{\hat{n}},\nu)\Re[\exp{i(\varphi_{k} +
    \frac{2\pi}{\lambda}x_{jk})}]J(\nu) d\vec{\hat{n}}d\nu,
\end{equation}
\noindent where $J(\nu)$ is a bandpass weighting. There is an
integration over detector area as well in eq.(\ref{seven4}), which
will be implicit until \S\ref{irf2}, where we explore the effect of
the detector area and make a suitable and practical approximation .
Notice that $x_{jk}$ does not depend on the direction
$\vec{\hat{n}}$. For now, we make the (crude) approximation that
$\lambda\equiv\lambda_{0}$ (the central wavelength), so that this `internal
phase factor' may be taken out of the integral over $\nu$ as well. As discussed in \S\ref{prelims} and \ref{subband1}, this
crude approximation need not be made. We will instead choose to view
the visibilities as `averages' over a certain `sub-bandwidth', in
which case this `internal phase' factor is an average over this
`sub-bandwidth'. We will discuss a more efficient way to deal with
this issue in \S\ref{prelims} and \ref{subband1}. For now, Eq.(\ref{seven4}) reads
\begin{eqnarray}
&&\mathcal{O}_{jk} =
  \Re\left[\exp(i\frac{2\pi}{\lambda_{0}}x_{jk})\times\right.\nonumber\\
&&\left.\int\int G\left(\vec{\hat{n}}\right)I(\vec{\hat{n}},\nu)\exp\left[i(\varphi_{k})\right] J(\nu)d\nu d\vec{\hat{n}}\right]\label{seven70}\\
&=&\Re\left[\exp(i\frac{2\pi}{\lambda_{0}}x_{jk})\times\right.\nonumber\\
&&\left.\underbrace{\int\int G\left(\vec{\hat{n}}\right)I(\vec{\hat{n}},\nu)\exp\left[i2\pi\vec{u}_k\cdot\vec{\hat{n}}\right] J(\nu)d\nu d\vec{\hat{n}}}\right].\label{seven7}
\end{eqnarray}
The quantity indicated by the underbrace in eq.(\ref{seven7}) is the {\em
Visibility} (defined in eq.(1)) from the $k$th baseline,
$V(\vec{u}_k)\equiv V_k$. We can now denote the `internal phase differences' 
$(2\pi/\lambda)\mathrm{ } x_{jk}$ as $\phi_{jk}$, so that
\begin{equation}
\mathcal{O}_{jk} = \Re\left[\exp(i\phi_{jk})V_k\right].
\end{equation}
In Appendix~\ref{fizeauimage}, we show that by integrating over the entire focal
plane, we can recover the image convolved with a ``dirty beam'', as in
radio interferometry.

In the next section, we consider the net signal from a single baseline
and describe how it can be split into `sub--bands.'  For the sake of
simplicity, the index $k$ corresponding to
baseline $k$ is dropped, and all equations in \S\ref{section3} hold for
every baseline.

\section{Spectral information from an interferometer using a Fizeau approach}
\label{section3}
%
%
\subsection{Preliminaries}
\label{prelims}
The output measured at the detectors in a Fizeau interferometer contains the following
phase information integrated over the entire bandwidth:
\begin{enumerate}
\item\label{irf11} phase introduced because of the path difference between any two
  rays that arrive from the same part of the sky on the two
  outward-facing antennas that make up a baseline; and
\item\label{irf12} phase introduced because of the path difference between any two
  rays that arrive from two different antennas on to the same point in
  the focal plane.
\end{enumerate}
The phase in point \ref{irf11} is due to the fact that we are
considering a radio interferometer, and so the visibility that we measure must, by
definition, include this phase. However, the phase in \ref{irf12}
above introduced by the beam combiner needs to be factored out to
recover visibility from each bolometer. If there were a way to
calculate the `internal phase' introduced by the beam combiner over
the whole bandwidth, then all we would need to do is to divide the
output at each point in the focal plane by this `internal phase', and we would get visibility directly\footnote{A range of values of $\nu$ will produce a range of $\ell$'s, or
a band in $\ell$-space (as discussed in \S\ref{context}, see eqs.(\ref{fizeau01},\ref{fizeau102})). A finite-bandwidth interferometer thus
measures what is called a `bandpower' instead of a single value of
the power spectrum at one value of $\ell$. But the power spectrum is
just the variance of the visibilities for a circle (ring) in the u--v
plane \citep[see][Fig.~5.8, Figs.~1\&2~respectively]{2007PhDT........31M,2010A&A...514A..37C}. And so we get different bandpowers for the same baseline and
orientation but for different frequencies.}.
\subsection{`Sub--band Splitting'}
\label{subband1}
We can think of the effect of the instrument on the visibilities
in the following way. Let us divide the entire bandwidth of the
instrument into $m$ sub-bands and let $\nu_{1}$, $\nu_{2}$...$\nu_m$ be the
centre-frequencies of each one.  Then for one baseline, one
orientation, and one detector position, these will correspond to
visibilities $V_{1}$, $V_{2}$...$V_{m}$ and to phase differences
$\phi_{j1}$, $\phi_{j2}$...$\phi_{jm}$ (where $j$ represents the
detector). If we represent the output at the $j$th detector as ${\mathcal
  O}_{j}$ then we get (as in the previous section, but dropping the index
$k$ for baseline, since we're considering just one baseline):
\begin{equation}
{\mathcal O}_{j} = \sum_{\alpha=1}^{m}\Re\left[V_{\alpha}\exp{i\phi_{j\alpha}}\right].
\end{equation}
Given just one detector, it is impossible to extract every
$V_{\alpha}$ for every sub-band, even though we know precisely what
the $\phi_{j\alpha}$'s are. However, if we have $m$ detectors, then we
can easily write the following system of equations for each baseline:
\begin{eqnarray}
{\mathcal O}_{1} &=& \Re[V_{1}\exp{i\phi_{11}} + V_{2}\exp{i\phi_{12}} + \cdots + V_{m}\exp{i\phi_{1m}}]\nonumber\\
{\mathcal O}_{2} &=& \Re[V_{1}\exp{i\phi_{21}} + V_{2}\exp{i\phi_{22}} +\cdots + V_{m}\exp{i\phi_{2m}}]\nonumber\\
\ldots\nonumber\\
{\mathcal O}_{m} &=& \Re[V_{1}\exp{i\phi_{m1}} + V_{2}\exp{i\phi_{m2}} +\cdots + V_{m}\exp{i\phi_{mm}}].
\label{irf102}
\end{eqnarray}
Now, as discussed in the preceding section and in Fig.(1), we can
apply a unique phase shift to each baseline. If we denote this phase
shift by $\Delta\varphi$ (it is understood that
$\Delta\varphi=\Delta\varphi(t)$) and the output after applying the
phase shift as ${\mathcal O}^{\prime}_{1}\ldots {\mathcal O}^{\prime}_{m}$, then we get another set of $m$
equations:
\begin{eqnarray}
{\mathcal O}^{\prime}_{1} &=& \Re[V_{1}\exp{i(\phi_{11}+\Delta\varphi)} + \cdots + V_{m}\exp{i(\phi_{1m}+\Delta\varphi)}]\nonumber\\
{\mathcal O}^{\prime}_{2} &=& \Re[V_{1}\exp{i(\phi_{21}+\Delta\varphi)} + \cdots + V_{m}\exp{i(\phi_{2m}+\Delta\varphi)}]\nonumber\\
\ldots\nonumber\\
{\mathcal O}^{\prime}_{m} &=& \Re[V_{1}\exp{i(\phi_{m1}+\Delta\varphi)} + \cdots + V_{m}\exp{i(\phi_{mm}+\Delta\varphi)}].
\label{irf1021}
\end{eqnarray}
This is a system of $2m$ equations with $2m$ unknowns - $\Re[V_1]$, $\Re[V_2]
\ldots \Re[V_m] \ldots \Im[V_1]$, $\Im[V_2] \ldots \Im[V_m]$, and so we can solve for the values for each one of these
`sub-band visibilities'. The beam combiner thus achieves far more
than just separating the real and imaginary parts of visibilities. 

\subsection{Effect of finite detector size}
\label{irf2}
We mentioned in \S\ref{prelims} that if the `internal phase' is known,
sub-band visibilities may be computed/estimated. However, calculating this internal phase is not easy, since
integration over the bandwidth complicates the calculation, as seen in
\S\ref{fizeau_interferometer}, eq.(\ref{seven7}), where the `internal
phase' factor was taken out of the integral to define a visibility
using a crude approximation \citep[see][\S2.2]{2010A&A...514A..37C}.
The internal phase differences in the Fizeau interferometer create
fringe patterns on the focal plane. %
This is exactly the same as saying that the
 visibilities in each sub-band are modulated by a fringe which
  depends on baseline length. To extract these visibilities, we need
  to separate the fringes. In order to do so, we need to realize that what we observe
at every detector is the visibility on the sky times the fringe summed over the area of the detector as well as bandwidth.

The fringe pattern is different for every frequency in the
bandwidth. Visibility is \emph{also} different for different
frequencies. Since we can compute the fringe corresponding to each
frequency, it is also straightforward to sum up these fringes over a
small `sub-band' over the area of a single detector. The output at
each detector from a baseline is known, and this can be written as a
sum over the product of visibility for a `sub-band' and the fringe
for the respective `sub-band'. This system of equations can be solved
for each baseline to yield the sub-band visibilities. We have
demonstrated this in \S\ref{subband1}. Effectively, this amounts to not using the crude
approximation in \S\ref{fizeau_interferometer} eq.(\ref{seven7}) and
instead writing the output as a product of a sub-band visibility and
a fringe. 

In the discussion so far, we have assumed that the collecting area of each detector is
negligible, and we completely ignored the effect of the fringe
pattern. Let us account for these effects in the following way. Let
$A$ be the effective collecting area of each detector. Let
$f\left(\vec{x},\nu_{\alpha}\right)$ be the value of the fringe
pattern (see discussion at the beginning of this section) at a point on the focal plane $\vec{x}$ and in a frequency
sub-band marked by $\alpha$. Then, equations (\ref{irf102},\ref{irf1021}) become
\begin{eqnarray}
{\mathcal O}_{j} &=& \sum_{\alpha=1}^{m}\Re\left[\int V_{\alpha}\exp{i\phi_{j\alpha}\left(\vec{x}\right)}f\left(\vec{x},\nu_{\alpha}\right)d^2\vec{x}\right]\nonumber\\
{\mathcal O}^{\prime}_{j} &=& \sum_{\alpha=1}^{m}\Re\left[\int V_{\alpha}\exp{i(\phi_{j\alpha}\left(\vec{x}\right)+\Delta\varphi)}f\left(\vec{x},\nu_{\alpha}\right)d^2\vec{x}\right]\label{irf103}
\end{eqnarray}
where it is understood that integration is done over the area of the
detector.

This leaves us with an issue - that of deconvolving the
$V$'s from the integrals. However, if the area of the detector is
small compared to the width of fringes, then we can assume that the
phase differences remain roughly constant over the collecting area of
one detector, so that we may write 
\begin{eqnarray}
{\mathcal O}_{j} &=& A\sum_{\alpha=1}^{m}\Re\left[V_{\alpha}\exp{i\phi_{j\alpha}\left(\vec{x}\right)}F\left(\vec{x},\nu_{\alpha}\right)\right]\nonumber\\
{\mathcal O}^{\prime}_{j} &=& A\sum_{\alpha=1}^{m}\Re\left[V_{\alpha}\exp{i(\phi_{j\alpha}\left(\vec{x}\right)+\Delta\varphi)}F\left(\vec{x},\nu_{\alpha}\right)\right]\label{irf104}
\end{eqnarray}
where $F\left(\vec{x},\nu_{\alpha}\right)$ represents an ``average''
value of the fringe pattern. 

Equations (\ref{irf104}) again have $2m$ variables and can be solved to
get $2m$ quantities: the real and imaginary parts of $m$ visibilities over the bandwidth.

\emph{Application to CMB cosmology: }The QUBIC collaboration is implementing the technique described in
this paper \citep{2010A&A...514A..37C}.

\section{Conclusions}
\begin{enumerate}
\item The Fizeau system makes it possible to recover spectral information without the need for filters.
\item The Fizeau system acts naturally as an imager.
\end{enumerate}
In addition, by introducing phase modulators discussed in \citep{2009MNRAS.393..531H,2009A&A...497..963C}, we can measure visibilities for all baselines in a Fizeau system.

\begin{figure}
  \begin{center}
    \includegraphics*[height=7cm, angle=0]{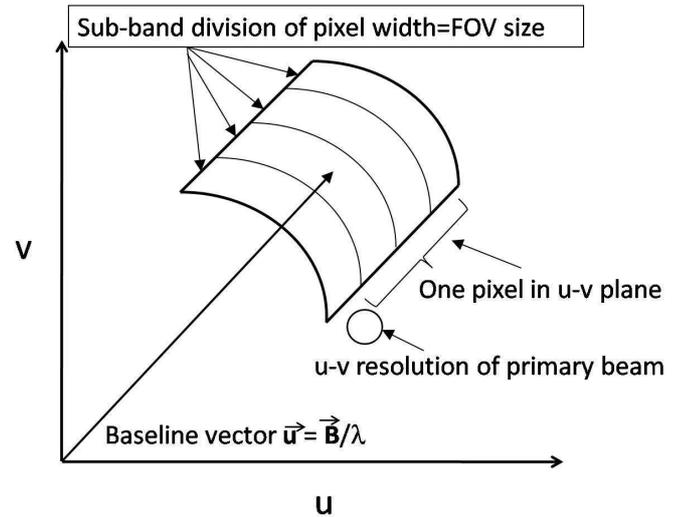}\\
    \caption{The u--v plane coverage of one baseline of an interferometer for a single pointing in a single baseline
    orientation angle. Radial spread in a single pixel in the u--v
    plane due to bandwidth is shown. Resolution in the u--v plane is
    determined by the primary beam or size of field-of-view. The
    minimum size of each sub--band is also determined by this
    resolution, as shown.}
    \label{302}
  \end{center}
\end{figure}

\begin{figure}
  \begin{center}
    \includegraphics*[height=10cm, angle=0]{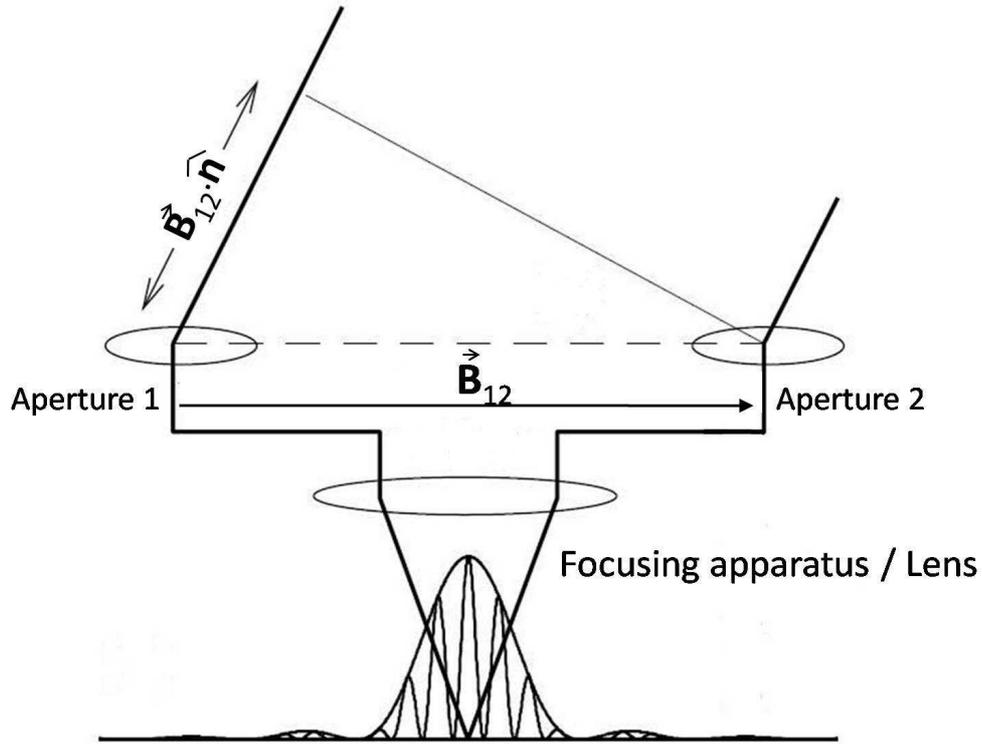}\\
    \caption{A Fizeau adding interferometer used as a beam combiner for long-baseline optical interferometry (figure reproduced
      from \cite{2000plbs.conf...31T}).  The baseline $\vec{B_{12}}$ is the separation of the apertures, which can be widely spaced, and determines the angular resolution of the instrument.  The Fizeau beam combiner has a different and much smaller baseline length (not labeled).  The`external' phase differences are marked $\vec{B_{12}}\cdot\vec{\hat{n}}$.}
    \label{youngs}
  \end{center}
\end{figure}

\begin{figure}
  \begin{center}
    \includegraphics*[height=10cm, angle=0]{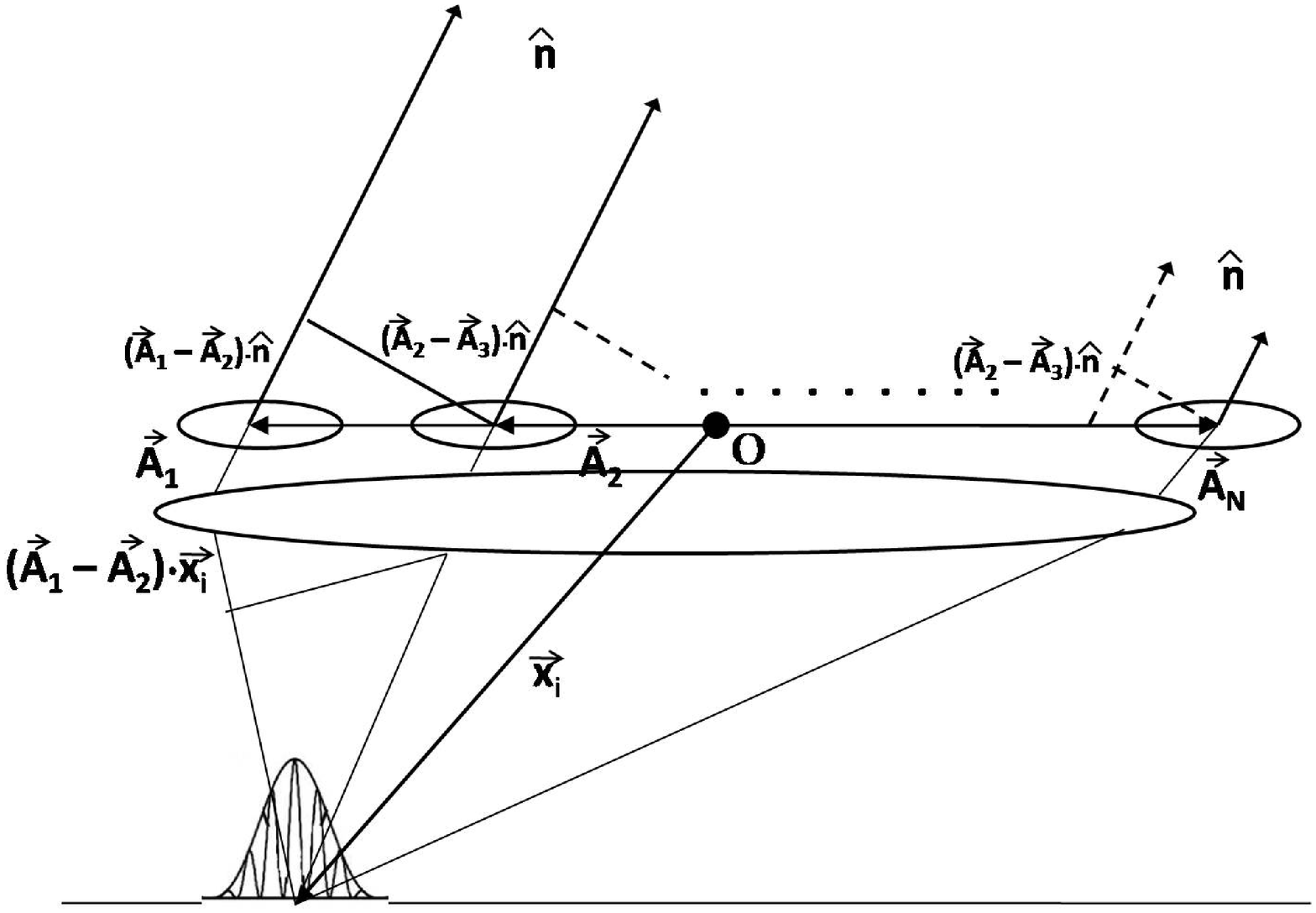}\\
    \caption{A simple 1-d Fizeau system, similar to a figure in
     \cite{2000plbs.conf...31T} .  {\bf O} denotes the origin of the co-ordinate
      system, and is the center of the observation plane. $\vec{x}_i$'s
      denote positions of detectors on the focal plane, and $\vec{A}_{l}$'s
      denote positions of apertures, where
      $l\in \left[1\ldots N\right]$. Notice that there are two
      sets of phase differences, marked $(\vec{A}_p-\vec{A}_l)\cdot\vec{\hat{n}}\equiv\vec{B}_{k}\cdot\vec{\hat{n}}$ and $x_{jk}(\vec{x})=\vec{x}_j\cdot(\vec{A}_p-\vec{A}_l)$ for
      external and internal phase differences respectively. The geometrical variation of $x_{jk}(\vec{x})=\vec{x}_j\cdot(\vec{A}_p-\vec{A}_l)$ allows
      ``sub--band splitting''.}
    \label{fizeau1d}
  \end{center}
\end{figure}

While it is possible to divide the bandwidth into many different
sub-bandwidths, it isn't possible to do this indefinitely. The beam for a
single antenna determines the FOV of the instrument and limits the
resolution in the u--v plane, as shown in Fig.(\ref{302}).

%
It is also possible to operate the interferometer \emph{simultaneously}
as an imager. The additional modulation mentioned above opens up a range of
possibilities, including the simultaneous measurement of visibilities
and images. This is described in Appendix~\ref{fizeauimage}.

In conclusion, the Fizeau system introduced here is potentially powerful tool
for astrophysics: it could allow the recovery of more information than is
possible with traditional interferometers or imagers and does not need
significantly more resources to build. Its application in CMB
cosmology is straightforward and can be demonstrated in future version
of QUBIC \citep{2010EAS....40..399H}.

\begin{acknowledgements}
We thank the members of the MBI and QUBIC collaborations for many fruitful discussions on bolometric interferometry.
\end{acknowledgements}

\bibliographystyle{aa}
\bibliography{Adding_Interferometry}

\begin{thebibliography}{26}
\expandafter\ifx\csname natexlab\endcsname\relax\def\natexlab#1{#1}\fi

\bibitem[{{Baumann} {et~al.}(2009){Baumann}, {Jackson}, {Adshead}, {Amblard},
  {Ashoorioon}, {Bartolo}, {Bean}, {Beltr{\'a}n}, {de Bernardis}, {Bird},
  {Chen}, {Chung}, {Colombo}, {Cooray}, {Creminelli}, {Dodelson}, {Dunkley},
  {Dvorkin}, {Easther}, {Finelli}, {Flauger}, {Hertzberg}, {Jones-Smith},
  {Kachru}, {Kadota}, {Khoury}, {Kinney}, {Komatsu}, {Krauss}, {Lesgourgues},
  {Liddle}, {Liguori}, {Lim}, {Linde}, {Matarrese}, {Mathur}, {McAllister},
  {Melchiorri}, {Nicolis}, {Pagano}, {Peiris}, {Peloso}, {Pogosian},
  {Pierpaoli}, {Riotto}, {Seljak}, {Senatore}, {Shandera}, {Silverstein},
  {Smith}, {Vaudrevange}, {Verde}, {Wandelt}, {Wands}, {Watson}, {Wyman},
  {Yadav}, {Valkenburg}, \& {Zaldarriaga}}]{2009AIPC.1141...10B}
{Baumann}, D., {Jackson}, M.~G., {Adshead}, P., {et~al.} 2009, in American
  Institute of Physics Conference Series, Vol. 1141, American Institute of
  Physics Conference Series, ed. {S.~Dodelson, D.~Baumann, A.~Cooray,
  J.~Dunkley, A.~Fraisse, M.~G.~Jackson, A.~Kogut, L.~Krauss, M.~Zaldarriaga,
  \& K.~Smith }, 10--120

\bibitem[{{Bock} {et~al.}(2009){Bock}, {Aljabri}, {Amblard}, {Baumann},
  {Betoule}, {Chui}, {Colombo}, {Cooray}, {Crumb}, {Day}, {Dickinson},
  {Dowell}, {Dragovan}, {Golwala}, {Gorski}, {Hanany}, {Holmes}, {Irwin},
  {Johnson}, {Keating}, {Kuo}, {Lee}, {Lange}, {Lawrence}, {Meyer}, {Miller},
  {Nguyen}, {Pierpaoli}, {Ponthieu}, {Puget}, {Raab}, {Richards}, {Satter},
  {Seiffert}, {Shimon}, {Tran}, {Williams}, \&
  {Zmuidzinas}}]{2009arXiv0906.1188B}
{Bock}, J., {Aljabri}, A., {Amblard}, A., {et~al.} 2009, ArXiv e-prints

\bibitem[{{Bock} {et~al.}(2006){Bock}, {Church}, {Devlin}, {Hinshaw}, {Lange},
  {Lee}, {Page}, {Partridge}, {Ruhl}, {Tegmark}, {Timbie}, {Weiss}, {Winstein},
  \& {Zaldarriaga}}]{2006astro.ph..4101B}
{Bock}, J., {Church}, S., {Devlin}, M., {et~al.} 2006, ArXiv Astrophysics
  e-prints

\bibitem[{{Bunn}(2007)}]{2007PhRvD..75h3517B}
{Bunn}, E.~F. 2007, Phys. ~ Rev. ~D, 75, 083517

\bibitem[{{Charlassier} {et~al.}(2010){Charlassier}, {Bunn}, {Hamilton},
  {Kaplan}, \& {Malu}}]{2010A&A...514A..37C}
{Charlassier}, R., {Bunn}, E.~F., {Hamilton}, J., {Kaplan}, J., \& {Malu}, S.
  2010, \aap, 514, A37+

\bibitem[{{Charlassier} {et~al.}(2009){Charlassier}, {Hamilton}, {Br{\'e}elle},
  {Ghribi}, {Giraud-H{\'e}raud}, {Kaplan}, {Piat}, \&
  {Pr{\^e}le}}]{2009A&A...497..963C}
{Charlassier}, R., {Hamilton}, J., {Br{\'e}elle}, {\'E}., {et~al.} 2009, \aap,
  497, 963

\bibitem[{{Charlassier} {et~al.}(2008){Charlassier}, {Hamilton}, {Br{\'e}elle},
  {Ghribi}, {Giraud-H{\'e}raud}, {Kaplan}, {Piat}, \&
  {Pr{\^e}le}}]{2008arXiv0806.0380C}
{Charlassier}, R., {Hamilton}, J.~C., {Br{\'e}elle}, {\'E}., {et~al.} 2008,
  ArXiv e-prints, 806

\bibitem[{{Chiang} {et~al.}(2010){Chiang}, {Ade}, {Barkats}, {Battle},
  {Bierman}, {Bock}, {Dowell}, {Duband}, {Hivon}, {Holzapfel}, {Hristov},
  {Jones}, {Keating}, {Kovac}, {Kuo}, {Lange}, {Leitch}, {Mason}, {Matsumura},
  {Nguyen}, {Ponthieu}, {Pryke}, {Richter}, {Rocha}, {Sheehy}, {Takahashi},
  {Tolan}, \& {Yoon}}]{2010ApJ...711.1123C}
{Chiang}, H.~C., {Ade}, P.~A.~R., {Barkats}, D., {et~al.} 2010, \apj, 711, 1123

\bibitem[{{Gupta} {et~al.}(2010){Gupta}, {Ade}, {Bock}, {Bowden}, {Brown},
  {Cahill}, {Castro}, {Church}, {Culverhouse}, {Friedman}, {Ganga}, {Gear},
  {Hinderks}, {Kovac}, {Lange}, {Leitch}, {Melhuish}, {Memari}, {Murphy},
  {Orlando}, {O'Sullivan}, {Piccirillo}, {Pryke}, {Rajguru}, {Rusholme},
  {Schwarz}, {Taylor}, {Thompson}, {Turner}, {Wu}, {Zemcov}, \& {QUaD
  Collaboration}}]{2010ApJ...716.1040G}
{Gupta}, S., {Ade}, P., {Bock}, J., {et~al.} 2010, \apj, 716, 1040

\bibitem[{{Hamilton} \& {Charlassier}(2010)}]{2010EAS....40..399H}
{Hamilton}, J. \& {Charlassier}, R. 2010, in EAS Publications Series, Vol.~40,
  EAS Publications Series, ed. {L.~Spinoglio \& N.~Epchtein}, 399--404

\bibitem[{{Hamilton} {et~al.}(2008){Hamilton}, {Charlassier}, {Cressiot},
  {Kaplan}, {Piat}, \& {Rosset}}]{2008arXiv0807.0438H}
{Hamilton}, J.~C., {Charlassier}, R., {Cressiot}, C., {et~al.} 2008, ArXiv
  e-prints, 807

\bibitem[{{Hinderks} {et~al.}(2009){Hinderks}, {Ade}, {Bock}, {Bowden},
  {Brown}, {Cahill}, {Carlstrom}, {Castro}, {Church}, {Culverhouse},
  {Friedman}, {Ganga}, {Gear}, {Gupta}, {Harris}, {Haynes}, {Keating}, {Kovac},
  {Kirby}, {Lange}, {Leitch}, {Mallie}, {Melhuish}, {Memari}, {Murphy},
  {Orlando}, {Schwarz}, {Sullivan}, {Piccirillo}, {Pryke}, {Rajguru},
  {Rusholme}, {Taylor}, {Thompson}, {Tucker}, {Turner}, {Wu}, \&
  {Zemcov}}]{2009ApJ...692.1221H}
{Hinderks}, J.~R., {Ade}, P., {Bock}, J., {et~al.} 2009, \apj, 692, 1221

\bibitem[{{Hu} {et~al.}(2003){Hu}, {Hedman}, \&
  {Zaldarriaga}}]{2003PhRvD..67d3004H}
{Hu}, W., {Hedman}, M.~M., \& {Zaldarriaga}, M. 2003, Phys. ~Rev. ~D, 67,
  043004

\bibitem[{{Hyland} {et~al.}(2009){Hyland}, {Follin}, \&
  {Bunn}}]{2009MNRAS.393..531H}
{Hyland}, P., {Follin}, B., \& {Bunn}, E.~F. 2009, \mnras, 393, 531

\bibitem[{{Komatsu} {et~al.}(2009){Komatsu}, {Dunkley}, {Nolta}, {Bennett},
  {Gold}, {Hinshaw}, {Jarosik}, {Larson}, {Limon}, {Page}, {Spergel},
  {Halpern}, {Hill}, {Kogut}, {Meyer}, {Tucker}, {Weiland}, {Wollack}, \&
  {Wright}}]{2009ApJS..180..330K}
{Komatsu}, E., {Dunkley}, J., {Nolta}, M.~R., {et~al.} 2009, \apjs, 180, 330

\bibitem[{{Kovac} {et~al.}(2002){Kovac}, {Leitch}, {Pryke}, {Carlstrom},
  {Halverson}, \& {Holzapfel}}]{2002Natur.420..772K}
{Kovac}, J.~M., {Leitch}, E.~M., {Pryke}, C., {et~al.} 2002, Nature, 420, 772

\bibitem[{Lawrence {et~al.}(2008)Lawrence, Church, Gaier, Lai, Ruf, \&
  Wollack}]{lawrence2008}
Lawrence, C.~R., Church, S., Gaier, T., {et~al.} 2008, J. Phys.: Conf. Series

\bibitem[{{Lin} \& {Wandelt}(2006)}]{2006APh....25..151L}
{Lin}, Y. \& {Wandelt}, B.~D. 2006, Astroparticle Physics, 25, 151

\bibitem[{{Malu}(2007)}]{2007PhDT........31M}
{Malu}, S.~S. 2007, PhD thesis, The University of Wisconsin - Madison

\bibitem[{{Monnier}(2003)}]{2003RPPh...66..789M}
{Monnier}, J.~D. 2003, Reports on Progress in Physics, 66, 789

\bibitem[{Rohlfs \& Wilson(2004)}]{Rohlfs&Wilson}
Rohlfs, K. \& Wilson, T.~L. 2004, Tools of Radio Astronomy (Springer)

\bibitem[{{Takahashi} {et~al.}(2010){Takahashi}, {Ade}, {Barkats}, {Battle},
  {Bierman}, {Bock}, {Chiang}, {Dowell}, {Duband}, {Hivon}, {Holzapfel},
  {Hristov}, {Jones}, {Keating}, {Kovac}, {Kuo}, {Lange}, {Leitch}, {Mason},
  {Matsumura}, {Nguyen}, {Ponthieu}, {Pryke}, {Richter}, {Rocha}, \&
  {Yoon}}]{2010ApJ...711.1141T}
{Takahashi}, Y.~D., {Ade}, P.~A.~R., {Barkats}, D., {et~al.} 2010, \apj, 711,
  1141

\bibitem[{{Timbie} {et~al.}(2006){Timbie}, {Tucker}, {Ade}, {Ali}, {Bierman},
  {Bunn}, {Calderon}, {Gault}, {Hyland}, {Keating}, {Kim}, {Korotkov}, {Malu},
  {Mauskopf}, {Murphy}, {O'Sullivan}, {Piccirillo}, \&
  {Wandelt}}]{2006NewAR..50..999T}
{Timbie}, P.~T., {Tucker}, G.~S., {Ade}, P.~A.~R., {et~al.} 2006, New Astronomy
  Review, 50, 999

\bibitem[{{Traub}(2000)}]{2000plbs.conf...31T}
{Traub}, W.~A. 2000, in Principles of Long Baseline Stellar Interferometry, ed.
  {P.~R.~Lawson}, 31--+

\bibitem[{{Tucker} {et~al.}(2008){Tucker}, {Korotkov}, {Gault}, {Hyland},
  {Malu}, {Timbie}, {Bunn}, {Keating}, {Bierman}, {O.Sullivan}, {Ade}, \&
  {Piccirillo}}]{2008SPIE}
{Tucker}, G.~S., {Korotkov}, A.~L., {Gault}, A.~C., {et~al.} 2008, in
  Millimeter and Submillimeter Detectors and Instrumentation for Astronomy IV.
  Edited by Zmuidzinas, Jonas; Holland, Wayne S.; Withington, Stafford; Duncan,
  William D. to appear in Proceedings of the SPIE (2008)., Presented at the
  Society of Photo-Optical Instrumentation Engineers (SPIE) Conference

\bibitem[{{Tuthill} {et~al.}(2000){Tuthill}, {Monnier}, {Danchi}, {Wishnow}, \&
  {Haniff}}]{2000PASP..112..555T}
{Tuthill}, P.~G., {Monnier}, J.~D., {Danchi}, W.~C., {Wishnow}, E.~H., \&
  {Haniff}, C.~A. 2000, \pasp, 112, 555

\end{thebibliography}
\onecolumn


\twocolumn
\appendix
\section{Imaging in a Fizeau system}
\label{fizeauimage}
In eq.(\ref{seven4}), if we now integrate over the focal-plane area, omitting the integral
over bandwidth, which is implicit (and remembering that $\varphi\equiv2\pi\vec{B}\cdot$/$\lambda$): 
\begin{equation}
\label{integration:5}
\mathcal{O} = \int\int\Re\left(G(\vec{\hat{n}})I(\vec{\hat{n}},\nu)\exp{\left[\mathrm{i}2\pi\frac{\vec{B}\cdot\vec{\hat{n}} - x_{ij}}{\lambda}\right]}\right)d\vec{\hat{n}} d^{2}\vec{x},
\end{equation}
\noindent where we have changed the sign on the `internal' phase
differences. This can be done without loss of generality, since the internal and external
phase differences are independent of each other.

Let us consider just one term in the expression $\Re\left(...\right)$:
\begin{equation}
\label{integration:6}
\mathcal{O} = \int\int I(\vec{\hat{n}},\nu)\exp{\left[\mathrm{i}2\pi\frac{\vec{B}\cdot\vec{\hat{n}}}{\lambda}\right]}d\vec{\hat{n}} \exp{\left[-\mathrm{i}2\pi\frac{x_{ij}}{\lambda}\right]}d^{2}\vec{x}
\end{equation}
If we include the effect of the primary beams of the inward--facing
antennas ($G(\vec{x})$) and adopt $I=I(\vec{\hat{n}},\nu)$,
\begin{equation}
\label{eight}
\mathcal{O} = \int
G\left(\vec{x}\right)\underbrace{\int
  G\left(\vec{\hat{n}}\right)I\exp{\left[\mathrm{i}2\pi\frac{\vec{B}\cdot\vec{\hat{n}}}{\lambda}\right]}d\vec{\hat{n}}}\mathrm{
}\exp{\left[-\mathrm{i}2\pi\frac{x_{ij}}{\lambda}\right]} d^{2}\vec{x}
\end{equation}
The quantity in underbrace is clearly a fourier transform, and
the expression can be written as
\begin{equation}
\label{nine}
\mathcal{O} = \int G\left(\vec{x}\right)\mathbf{\mathfrak{F}}\left(GI\right)\mathrm{ }\exp{\left[-\mathrm{i}2\pi\frac{x_{ij}}{\lambda}\right]} d^{2}\vec{x}
\end{equation}
If the distance from the inward-facing antennas to the focal
plane $\gg$ the collecting area for each bolometer,
\begin{equation}
\label{ten}
\mathcal{O} = \mathbf{\mathfrak{F}}^{-1}\left(G\mathbf{\mathfrak{F}}\left(GI\right)\right)
\end{equation}
The beam needs to be deconvolved from the
above expression in order to obtain an image from the instrument.

Now, eq(\ref{nine}) can be split up over the focal plane:
\begin{equation}
\label{eleven}
\mathcal{O} = \sum_{i=1}^{N} \int_{i}G\mathbf{\mathfrak{F}}\left(GI\right)\mathrm{
  }\exp{\left[-\mathrm{i}2\pi\frac{x_{ij}}{\lambda}\right]}d^2\vec{x}
\end{equation}
where $1\ldots N$ are labels for bolometers on the focal plane. 

Each of the bolometer outputs then represents a pixel in image
space. The total number of pixels depends on the resolution
of the instrument, and not the number of bolometers on the focal
plane. Therefore, if the number of bolometers on the focal plane are
{\bf greater} than the number of pixels in the image, we need to
``repixelize'' the image obtained, so that all pixels are independent
of each other.

In general, this is how the beam is convolved with the image on the
sky for the Fizeau beam combiner:
\begin{eqnarray}
\mathcal{O} &=&\mathbf{\mathfrak{F}}^{-1}\left(G\mathbf{\mathfrak{F}}\left(GI\right)\right)\label{nontrad1}\\
&=& \left[\left(\mathbf{\mathfrak{F}}^{-1}G\right)   \ast \left(GI\right)\right]\label{nontrad2}
\end{eqnarray}
In traditional interferometry, eq.(\ref{nontrad1}) would read
\begin{eqnarray}
\mathcal{O} &=&\mathbf{\mathfrak{F}}^{-1}\left(\mathbf{\mathfrak{F}}\left(GI\right)\right)\label{trad1}\\
&=&\mathbf{\mathfrak{F}}^{-1}\left(\mathbf{\mathfrak{F}}G\right)\ast\left(\mathbf{\mathfrak{F}}I\right)\label{trad2}\\
&\equiv & GI\label{trad3}
\end{eqnarray}
In eq.(\ref{trad2}), the u--v space beam $\mathbf{\mathfrak{F}}G$
needs to be multiplied by u--v coverage, which is a ``mask'', say
$\mathfrak{M}(u,v)$. Then, $\mathbf{\mathfrak{F}}^{-1}\left(\mathbf{\mathfrak{F}}G\times\mathfrak{M}(u,v)\right)$
is called the ``dirty beam'' in traditional interferometry. In
eq.(\ref{nontrad1}), the factor $\mathfrak{M}(u,v)$ is
included. Eq.(\ref{nontrad2}) thus tells us that the dirty beam for
the image produced by the Fizeau combiner is more involved than the
traditional interferometer dirty beam, but remains conceptually equivalent.

There are two assumptions inherent in the foregoing discussion:
\begin{enumerate}
\item The focal plane is large enough to receive most of the power
  from the inward-facing antennas
\item There are no ``blank'' areas on the focal plane for which the
  incident power is not absorbed by a bolometer
\end{enumerate}
It is possible to detect the correlated signal from a pair of antennas
as well. In order to separate every unique baseline, time--varying
phase--shifts can be applied to the signal at the base of the
skyward--facing antennas. The ``correlated signal'' from each pair of
antennas is simply the visibility from a baseline, with one crucial
difference: there is an ``internal phase'' added to every visibility
due to the relative positions of antennas on the observation plane and
detectors on the focal plane. These phase differences are
geometrical. Since the output from all $N$ antennas is incident on every detector, we can say that the output from
a single detector contains information about $N(N-1)/2$ visibilities. 


\end{document}